\documentclass[a4paper,11pt]{scrartcl}
\usepackage{ILD}
\usepackage{tagging}
\usetag{ILD}
\usepackage{subcaption}
\usepackage{hyperref}
\usepackage{wasysym}
\usepackage{amssymb}
\usepackage[backend=biber,style=numeric-comp,sorting=none,mcite=true,doi=false,subentry]{biblatex}
\usepackage{lineno}
\usepackage{slashed}
\usepackage{wrapfig,rotating}
\usepackage{pgfpages}
\usepackage{fancybox,graphicx}
\usepackage{graphbox}
\usepackage{epstopdf}
\AppendGraphicsExtensions{.eps.gz}
\title{
  Search for dark photons at future e$^+$e$^-$ colliders

}
\ildproc{phys}{2026}{009}
\localrep{DESY-26-042\\}
\date{\today}

\addauthor{
  Mikael Berggren

}
{\institute{1}}

\addinstitute{1}{
  Deutsches Elektronen-Synchrotron DESY,
Notkestr.~85, 22607~Hamburg, Germany

}
\onbehalfof{
    \tagged{POS}{On behalf of }\tagged{SPRINGER}{On behlf of }the ILD concept group

}

\abstract{
  In a class of theories, dark matter is explained by postulating the existence of a `dark sector',
which interacts gravitationally with ordinary matter. If this dark sector contains a U(1) symmetry,
and a corresponding `dark' photon ($A_{D}$) , it is natural to expect that this particle kineticly mix
with the ordinary photon, and hence become a `portal' through which the dark sector can be studied.
The strength of the mixing is given by a mixing parameter $(\epsilon)$. This
same parameter governs both the production and the decay of the $A_{D}$ back to SM
particles, and for values of  $\epsilon$ not already excluded, the signal would be
a quite small, and quite narrow resonance: If $\epsilon$ is large enough to
yield a detectable signal, its decay width will be smaller than the detector resolution, but so large
that the decay back to SM particles is prompt. For masses of the dark photon above the reach of
Belle II,  future high energy e$^+$e$^-$ colliders are ideal for searches for such a signal, due to the
low and well-known backgrounds, and the excellent momentum resolution and equally
excellent track-finding efficiency of the detectors at such colliders.
This contribution will discuss a study investigating the dependency of the limit on the mixing
parameter and the mass of the $A_{D}$ using the $A_{D}\rightarrow\mu^{+}\mu^{-}$ decay mode in
the presence of standard model background, using fully simulated signal and background events in
the ILD detector at the ILC Higgs factory. In addition, a more general discussion about the capabilities
expected for generic detectors at e$^+$e$^-$ colliders operating at other energies will be given.

}

\titlecomment{%
  Contributed to the Linear Collider Workshop (LCWS2025)\\
 20-24 October 2025\\
 Valencia,  Spain

}

\def\leqsim{\mathbin{\;\raise1pt\hbox{$<$}\kern-8pt\lower3pt\hbox{$\sim$}\;}}
\def\geqsim{\mathbin{\;\raise1pt\hbox{$>$}\kern-8pt\lower3pt\hbox{$\sim$}\;}}

\def\p#1{\mbox{$ \mbox{\bf p}_1                                         $}}

\newcommand{\eeto}    {\mbox{$ {\, \mathrm e}^+ {\mathrm e}^- \to             $}}

\newcommand{\GeV}     {\mbox{$ {\mathrm{GeV}}                              $}}

\newcommand{\ba}{\begin{array}}
\newcommand{\ea}{\end{array}}
\newcommand{\bc}{\begin{center}}
\newcommand{\ec}{\end{center}}
\newcommand{\be}{\begin{eqnarray}}
\newcommand{\eeq}{\end{eqnarray}}
\newcommand{\bes}{\begin{eqnarray*}}
\newcommand{\ees}{\end{eqnarray*}}
\newcommand{\Kz}{\ifmmode {\rm K^0_s} \else ${\rm K^0_s} $ \fi}
\newcommand{\Zz}{\ifmmode {\rm Z^0} \else ${\rm Z^0 } $ \fi}
\newcommand{\xxbar}{\ifmmode {\rm x\bar{x}} \else ${\rm x\bar{x}} $ \fi}
\newcommand{\rphi}{\ifmmode {\rm R\phi} \else ${\rm R\phi} $ \fi}

\def    \missEt      {\ifmmode{/\mkern-11mu E_t}\else{${/\mkern-11mu E_t}$}\fi}
\def    \missE       {\ifmmode{/\mkern-11mu E}\else{${/\mkern-11mu E}$}\fi}
\def    \missp       {\ifmmode{/\mkern-11mu p}\else{${/\mkern-11mu p}$}\fi}
\def    \misspt      {\ifmmode{/\mkern-11mu p_t}\else{${/\mkern-11mu p_t}$}\fi}

\begin{document}
\titlepage

  \section{Introduction}

Numerous cosmological and astronomical observations clearly confirm the existence of
  dark matter (DM), which accounts for 85\% of the matter in the universe.
  Weakly Interacting Massive Particles (WIMPs) continue
  to be the focus of observational efforts,
  but all such attempts have failed so far.
  This has led to another hypothesis that dark matter exists in a \textit{dark, hidden sector} in which
  its interaction with the visible sector is extremely weak.

Feebly interacting particles (FIPs) are one type of such models.
In these models, dark matter resides in the ``dark sector'', which is
neutral with respect to the Standard Model (SM) and interacts with the visible sector only via gravity.
However, it is thought that part of the dark sector interacts with the visible sector,
albeit in a very weak form.
Therefore, in these models,
  the reason BSMs have not yet been observed is not due to a lack of energy,
  but rather a lack of precision - whether in terms of  luminosity, sensitivity, background levels, 
or detector performance.

The mechanism that acts as a ``window'' into this dark region is known as the portal. 
There are many types of portals, and they 
are generally classified according to their composition into Higgs portals, fermion portals, pseudo-scalar 
portals and vector 
portals. 
These correspond to cases where the Higgs particle, a sterile neutrino, an axion-like particle (an ALPS), 
or a dark photon  
act as the mediator, respectively.
Here we will examine the last one.
\section{ The Vector Portal - Dark Photons, $A_D$}

    \begin{figure}[b]
      \begin{center}
       \subcaptionbox{}{
          \includegraphics [width=0.6\textwidth]{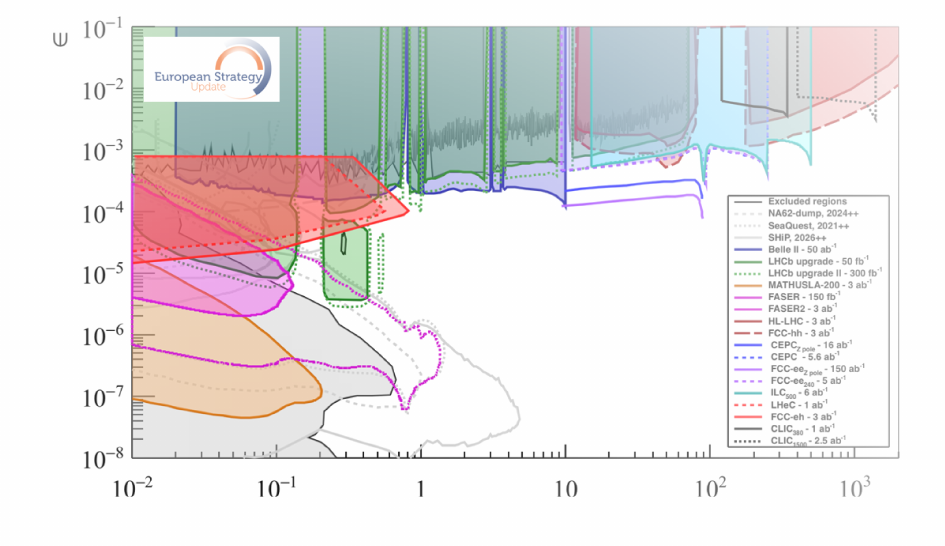}
        }
        \subcaptionbox{}{
          \includegraphics [width=0.35\textwidth]{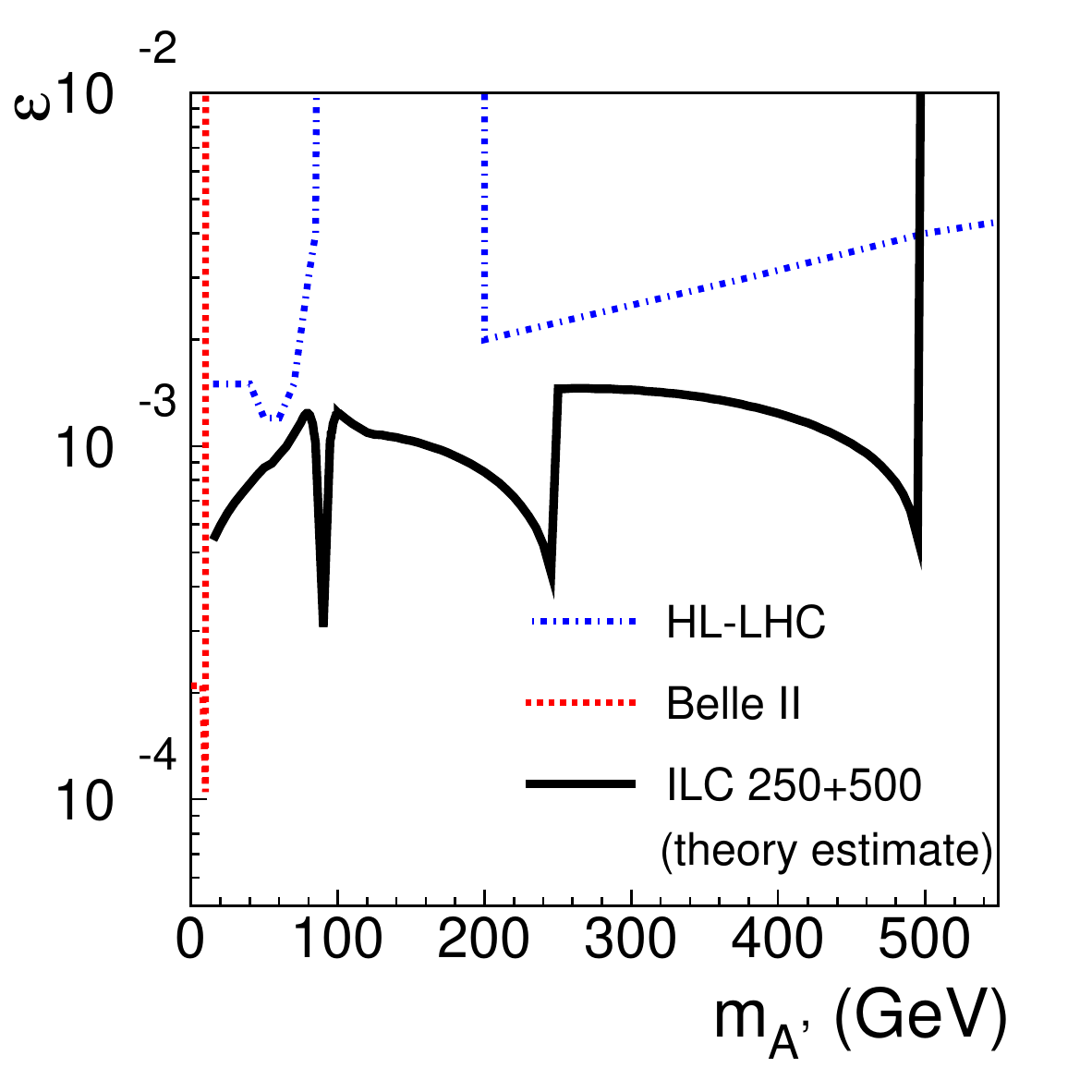}
        }
      \end{center}
      \caption{Expected Dark photon limits from 2019 EPPSU briefing-book~\cite{EuropeanStrategyforParticlePhysicsPreparatoryGroup:2019qin}. (a) Original figure, with all
        experiments considered and on a logarithmic mass-scale; (b) the same on a linear mass scale (up to the reach
        of the higgs factories), only showing the
        relevant experiments on this scale (Belle II, ILC 250 and HL-LHC).\label{fig:EPPSUdarkphoton}}
    \end{figure}
  Assume that there is a dark sector with a dark U(1) symmetry.
  The relevant part of the Lagrangian is \cite{Curtin:2014cca}:
    $$
\mathcal{L}_{gauge} = -\frac{1}{4} \,\hat B_{\mu\nu}\, \hat B^{\mu\nu} - \frac{1}{4} \,\hat Z_{D\mu\nu}\, \hat Z_D^{\mu\nu}  + 
\frac{1}{2}\,\frac{\epsilon}{\cos\theta_W} \,\hat Z_ {D\mu\nu}\,\hat B^{\mu\nu} 
+ \frac{1}{2}\, m_{D,0}^2\, \hat Z_D^\mu \, \hat Z_{D\mu}\,
$$
$\hat B$ is the ordinary U(1) field-strength tensor, and $\hat Z_D$ that of the
dark U(1).
The Dark Photon (the $A_D$) might mix with the photon by \textit{kinetic mixing}- the $\hat Z_ {D}\hat B$ term -
so that
  $
  \eeto A_D \rightarrow f\bar{f}
  $
  is possible.

The free mixing parameter, $\boldsymbol{\epsilon}$, must be small; otherwise, this process would have been observed by now.
      Few events will occur, but their decay will form a very narrow peak, or even a  displaced vertex.
    One must not forget that the dark photon itself is not the DM particle, as it is unstable;
      for this, another, stable particle is also required in the dark sector.

The current projections for Dark Photon limits from 
the 2019 update to the ``European Strategy for Particle Physics'' ~\cite{EuropeanStrategyforParticlePhysicsPreparatoryGroup:2019qin}
  are shown in Figure \ref{fig:EPPSUdarkphoton}(a), and on the linear mass scale, 
they are shown in Figure \ref{fig:EPPSUdarkphoton}(b).
  In the mass range up to $\sim$ 1 GeV, dark photons have long lifetimes and are likely to be detected more effectively in 
beam-dump experiments.
    At higher energies, colliding particle accelerators are required.
    Up to 10 GeV, the B-factories are the most powerful tool due to their extremely high luminosity.
    At higher energies, e$^+$e$^-$  up to the maximum energy provide the best sensitivity,
    and beyond that, hadron colliders show some effectiveness, but only if the coupling constant is quite high.
    It should be noted that the curves for the ILC (and CEPC, FCC-ee) are highly simplified theoretical estimates
    ~\cite{Karliner:2015tga}
       and, as we shall see later, these are
        clearly too optimistic.

\section{Dark Photons at  Higgs factories and beyond}

    \begin{figure}[t]
    \begin{center}
      \subcaptionbox{}{
        \includegraphics [width=0.45\textwidth]{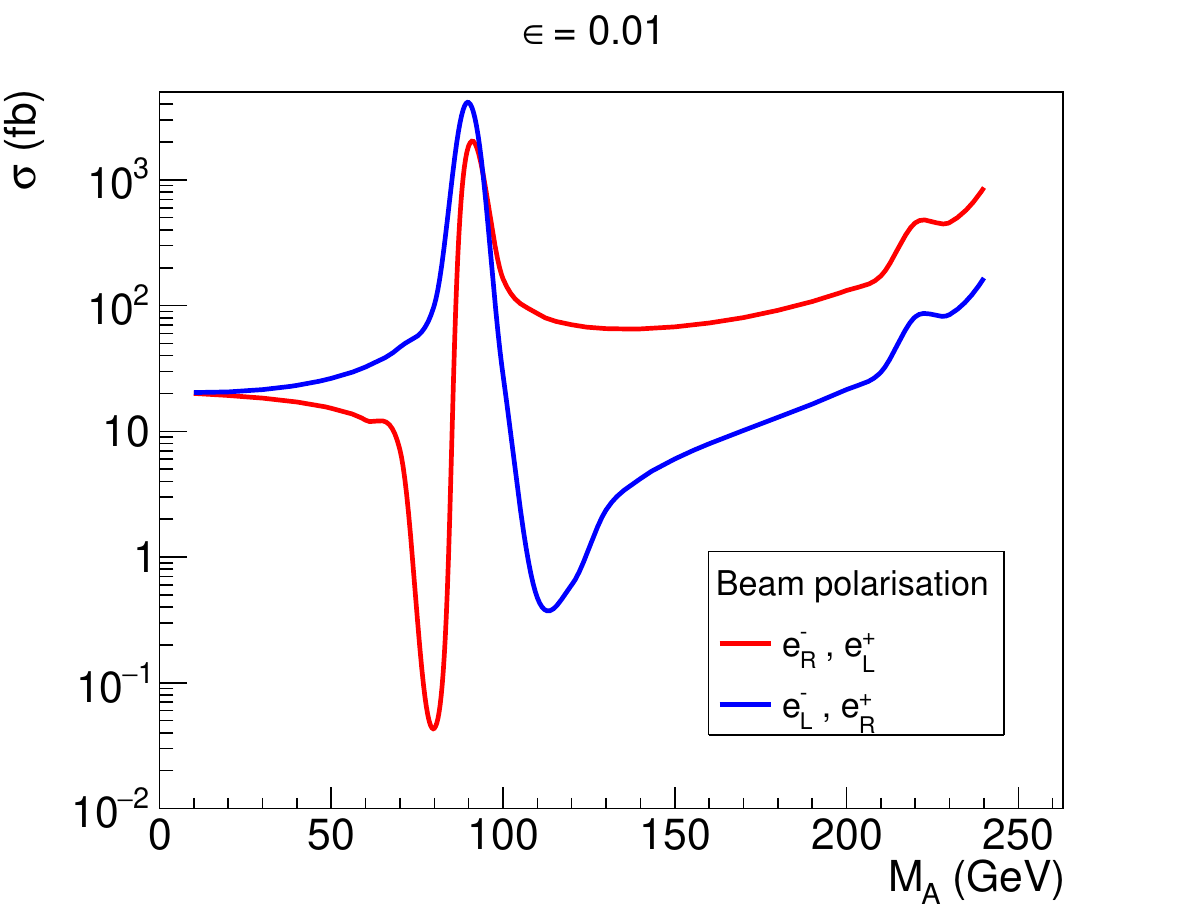}
      }%mass_brXxsect_RL_LR_isr_screen}
      \subcaptionbox{}{
        \includegraphics [width=0.45\textwidth]{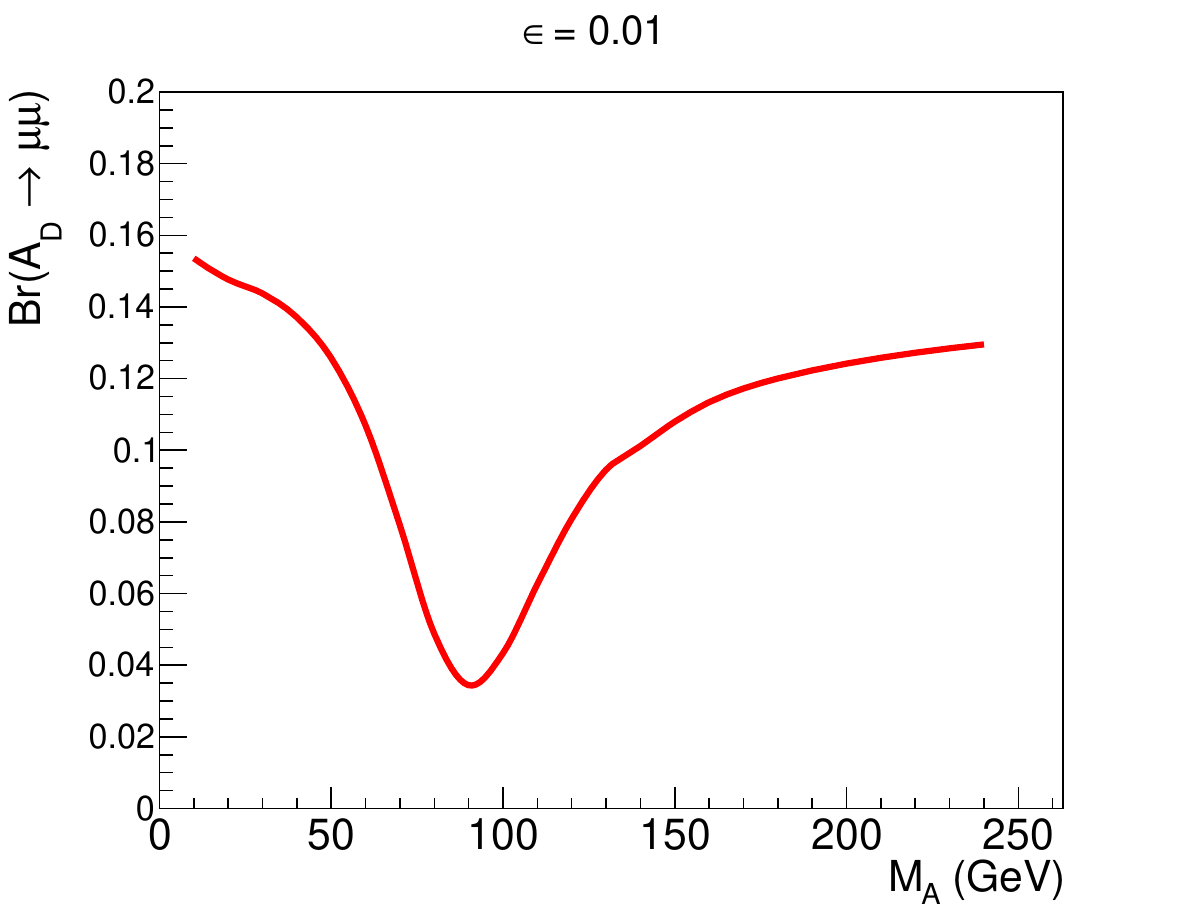}
      }

       \subcaptionbox{}{
        \includegraphics [width=0.45\textwidth]{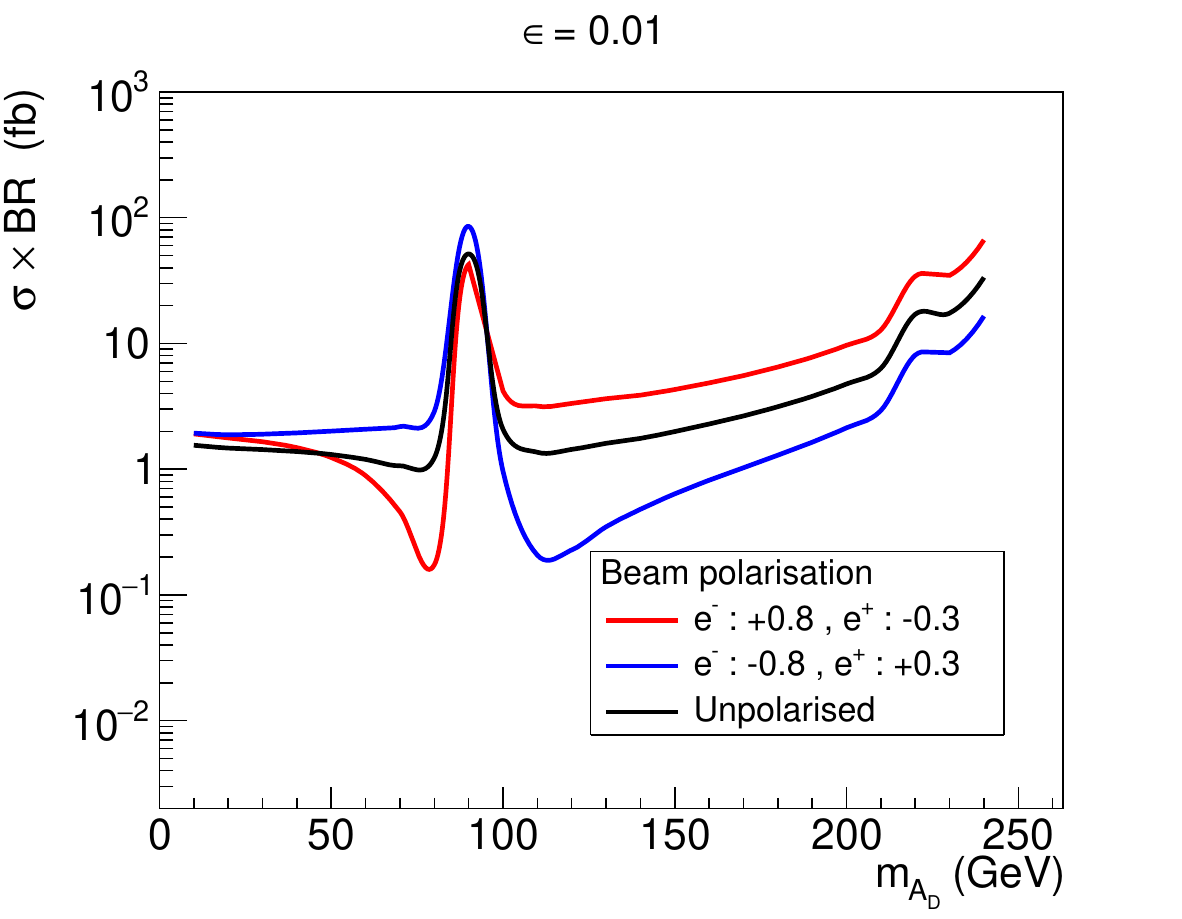}
      }%mass_brXxsect_RL_LR_isr_screen}
      \subcaptionbox{}{
        \includegraphics [width=0.45\textwidth]{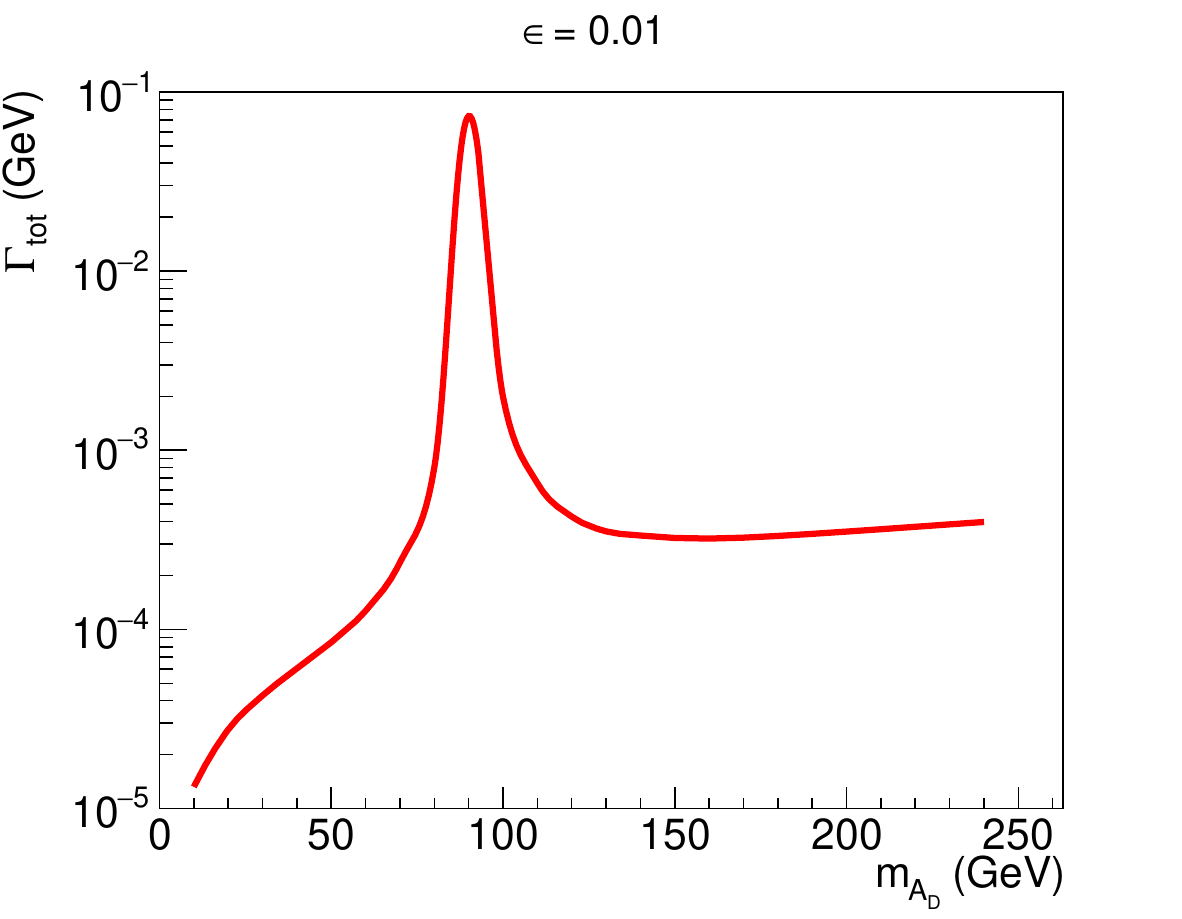}
      }
   \end{center}
    \caption{(a): Cross-section 
      of $\eeto \gamma_{ISR} A_D \rightarrow \mu^+ \mu^-  \gamma_{ISR}$, for the beam polarisations
      given in the legend; (b) Branching ratio of $A_D \rightarrow \mu^+ \mu^-$ 
      (c): Cross-section times branching ratio of $\eeto \gamma_{ISR} A_D \rightarrow \mu^+ \mu^-  \gamma_{ISR}$,
       for the beam polarisations
      given in the legend; (d) Total width of $A_D$.
      \label{fig:cross_sections_and_width}}
    \end{figure}
The proposed linear collider facilities acting as Higgs factories are the International Linear Collider (ILC)
~\cite{Behnke:2013xla},
the proposed Linear Collider Facility (LCF)~\cite{LinearCollider:2025lya},
the Cool Copper Collider (C3) ~\cite{Vernieri:2022fae},
and the Compact Linear Collider (CLIC) ~\cite{Brunner:2022usy}.
In addition to operating at the optimal energy for Higgs boson research,
these facilities are expected to benefit from energy upgrades that will enable them to reach TeV-scale energies.
 Circular colliders - the Circular Electron-Positron Collider
 (CEPC)~\cite{Gao:2022lew}
and
the Future Circular Collider (FCC-ee) ~\cite{Bernardi:2022hny}
focus more on low energies (although it is possible to increase the energy to $365$\,GeV). In particular,
very high luminosity can be achieved when operating at the Z-pole. 
Although these facilities, at their Higgs factory stage, only reach marginally higher than what LEP-II did,
a significant improvement in detection limits for the search for $A_D$ is expected:
They will have a luminosity at least 1,000 times higher  and -
in the case of linear machines - are characterised by
trigger-free operation and by having polarised beams.
Added to this is the benefit of 40 years' experience in detector development, and
indeed, numerous detector concepts have been proposed.
In this work, based on~\cite{Hosseini-Senvan:2024}, we have investigated the capabilities of one 
of these namely the  
International Large Detector concept 
(ILD)~\cite{ILDConceptGroup:2020sfq} at the ILC operating at 250 GeV.

The signal process for $A_D$ production in $e^+e^-$ collisions is
as follows: $\eeto \gamma_{ISR} A_D \rightarrow \mu^+ \mu^-  \gamma_{ISR}$, where
  the energy of the ISR is such that the recoil mass for the ISR is $m_{A_D}$.
To study this process, we generated events
  in accordance with the model presented in ~\cite{Curtin:2014cca}.
  We used the Unified Feynrules output file describing the model, provided by the authors, as input data for the event generator
  \texttt{Whizard} (ver. 3.0)~\cite{Kilian:2007gr}.
We see that both the production cross-section ($\sigma$) and the decay width ($\Gamma$) are proportional to $\epsilon^2$.
As a rough estimate, it seems likely that in every Higgs factory, $\sigma > \mathcal{O}$ (1 fb) can be reached.
As for the value of $\epsilon^2$ that yields such an order of magnitude for the cross-section times 
branching ratio
(see Figure \ref{fig:cross_sections_and_width}(c)),
it turns out that $\Gamma$ varies, depending on the mass of the dark photon, from $\mathcal{O}$(10 keV) to $\mathcal{O}$(10 MeV).
(See Figure \ref{fig:cross_sections_and_width}(d)).
This implies that the decay occurs instantaneously  with a c$\tau <$~1~nm, and that the width of the 
observed peak 
is determined not by the natural width but by the resolution of the detector
(see Figure \ref{fig:dimumass}(a)).

\section{Dark Photons in a real detector}

 \begin{figure}
   \begin{center}
      \subcaptionbox{}{
        \includegraphics [trim={0cm 0.0cm 0cm 0cm },clip,width=0.3\textwidth]{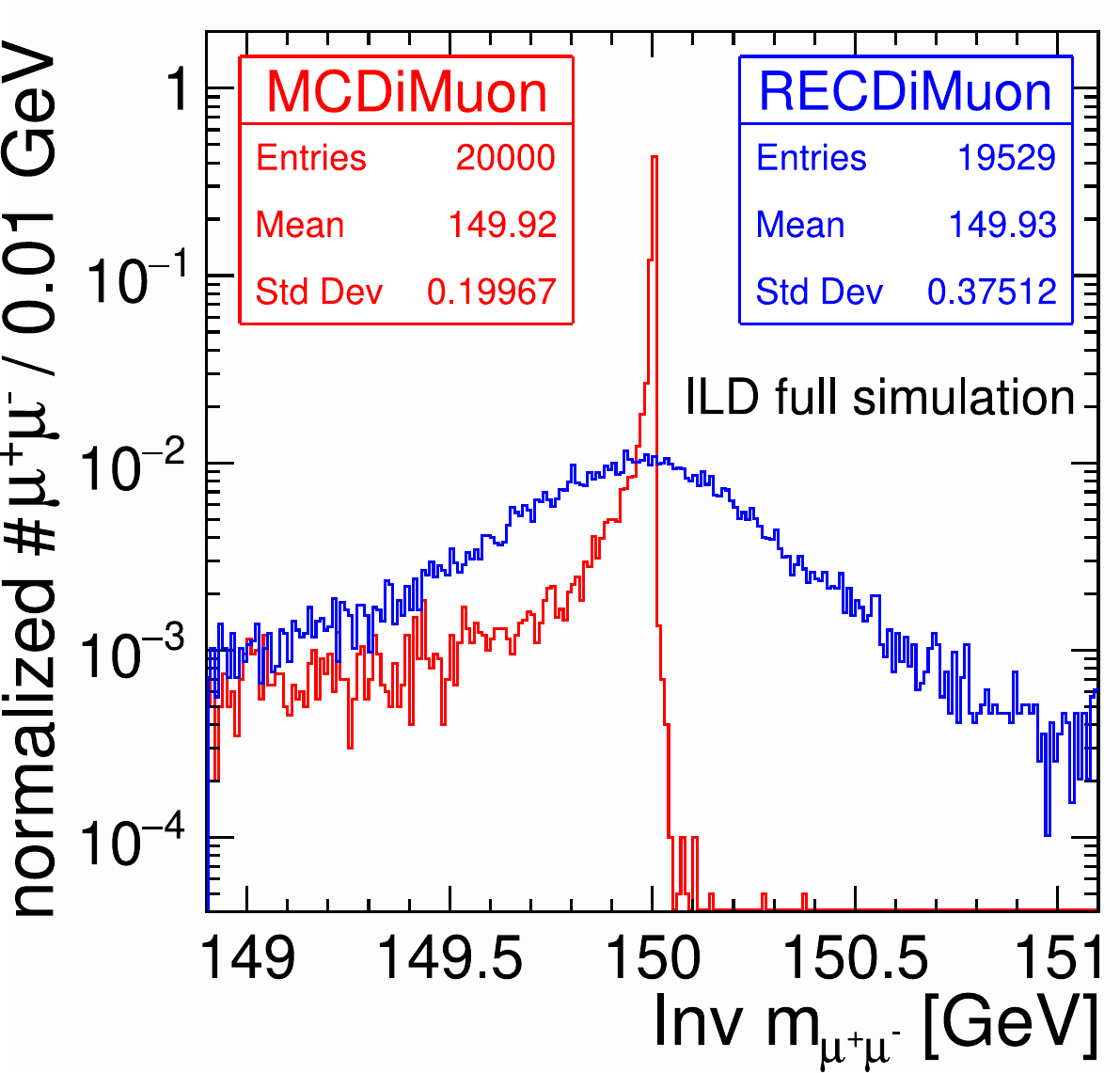}
      }
     \subcaptionbox{}{
       \includegraphics [width=0.3\textwidth]{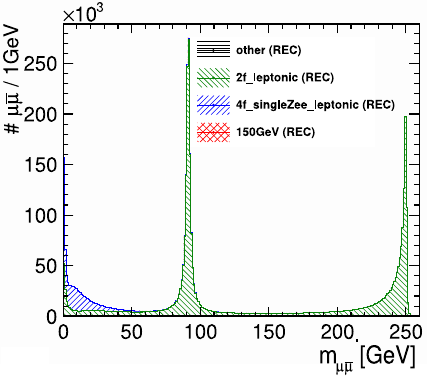}
     }
     \subcaptionbox{}{
       \includegraphics [trim={0cm 0.0cm 0cm 0cm },clip,width=0.3\textwidth]{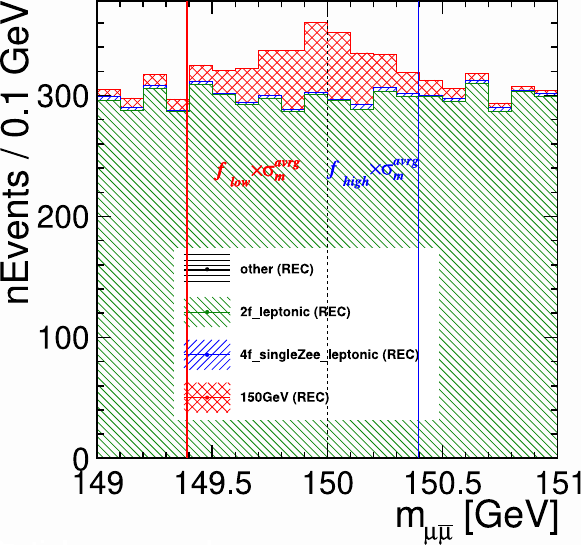}
     }
    \end{center}
    \caption{Di-muon mass distributions for $m_{A_D}$ = 150 GeV from full simulation of ILD: 
(a): Generated di-muon mass distribution (red), and reconstructed (blue), for a dark photn with mass 150 GeV.
(b): as (a), but also including all backgrounds; (c) is the same as (b) zoomed into the signal region. 
 \label{fig:dimumass}}
    \end{figure}
    \begin{figure}[b]
    \begin{center}
    \subcaptionbox{}{
        \includegraphics[trim={5.0cm 17.0cm 5.5cm 2cm },clip,width=0.45\textwidth]{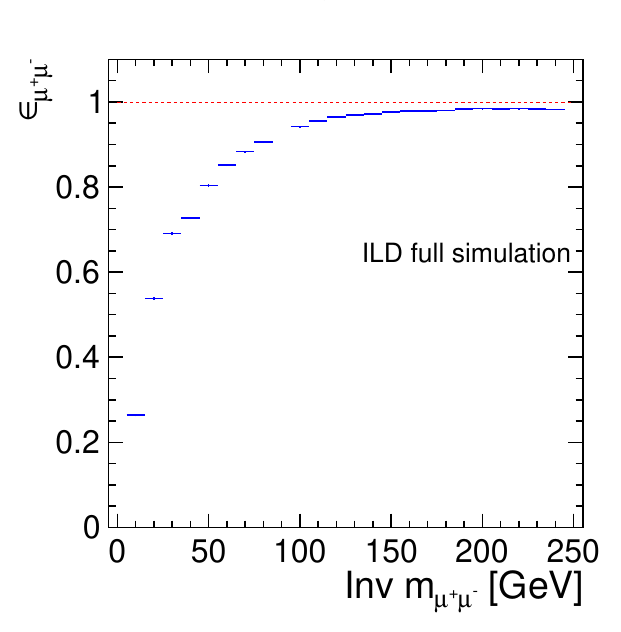}
    }
      \subcaptionbox{}{
        \includegraphics [trim={0cm 0cm 0cm 0cm },clip,width=0.48\textwidth]{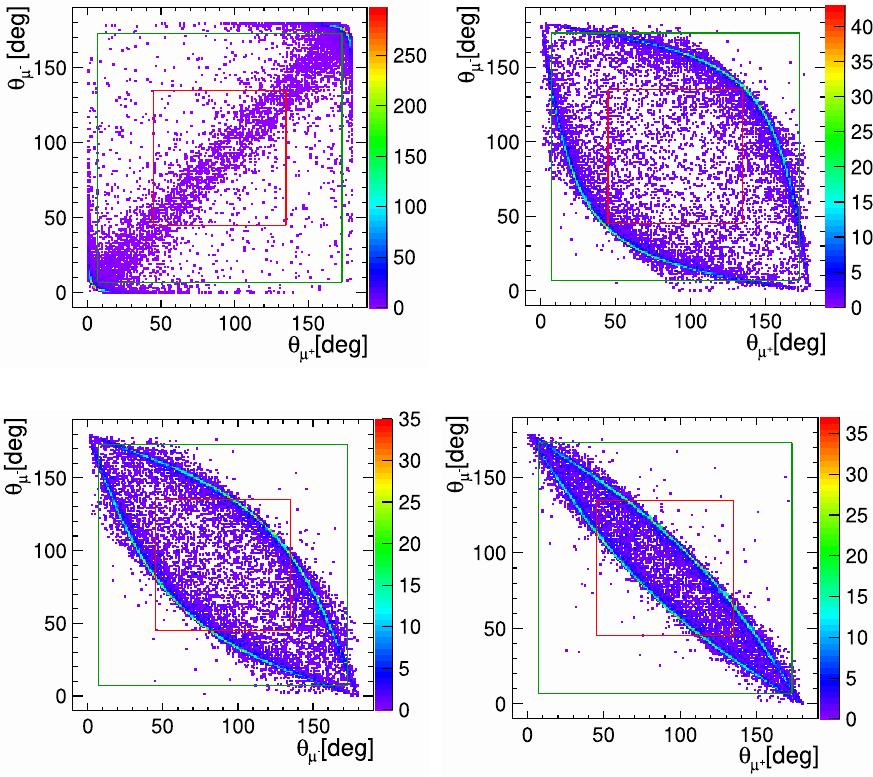}
      }
    \end{center}
    \caption{
      (a) The efficiency to find both muons from the decay of a  $A_D$ vs.  $m_{A_D}$;
      (b) The polar angle of the $\mu^-$ versus that of the $\mu^+$ of
      the generated  $\eeto \gamma_{ISR} A_D \rightarrow \mu^+ \mu^-  \gamma_{ISR}$ events,
      for $m_{A_D}$ = 10, 100, 150 and 200 GeV (clock-wise, from upper-left). The green square indicates the acceptance
      of the tracking system of ILD, and the red one indicates the coverage of the barrel tracking system;
\label{fig:effmu_and_thetamu}}
    \end{figure}
The generated events were then processed through ILD full simulation (\texttt{ddsim}~\cite{Petric:2017psf}), 
based entirely on \texttt{Geant4}, and full ILD reconstruction chain (\texttt{Marlin}~\cite{Gaede:2006pj}).
    In the analysis, the entire fully simulated SM background was taken into account~\cite{Berggren:2021sju}\footnote {
  $\eeto \mu^+\mu^- + ISR$ not only contributes to the background, but also %(Fig. ddda),
  t-channel processes involving undetected beam-remnant electrons, as well as those mistaken for ISR photons. 
These were also taken into account.}.
  The subsequent procedure involved
  only selecting  events containing two muons and, possibly, an isolated photon, but nothing else.
  In the sample of selected two-muon events, we search for an  arbitrarily small peak 
within the $m_{\mu\mu}$ distribution.
  Its width is determined by the detector resolution rather than the intrinsic width.
  Figure \ref{fig:dimumass}(b) shows the total mass spectrum, including all backgrounds.
    Figure \ref{fig:dimumass}(c) is an enlarged view of the signal region, corresponding in this case to a dark photon
    with a mass of 150 GeV.

Figure \ref{fig:effmu_and_thetamu}(a) shows the efficiency to detect the two muons as a function
of the dark photon mass. The detection efficiency is only about 25\% for $m_{A_D}$ = 10 GeV, but approaches
    100\% when $m_{A_D}$ is equal to or greater than 100 GeV.
This can be understood from Figure \ref{fig:effmu_and_thetamu}(b) which shows the angular distribution 
of the two muons for four different dark photon masses.
The green square marks
    the detection area of the track detectors and, as can be seen, a significant proportion of
    events are lost, especially for the smallest masses, for the simple reason that at least one of the muons lies at angles smaller than the acceptance of the track detectors.

    \begin{figure}[t]
    \begin{center}
      \subcaptionbox{}{
        \includegraphics [width=0.48\textwidth]{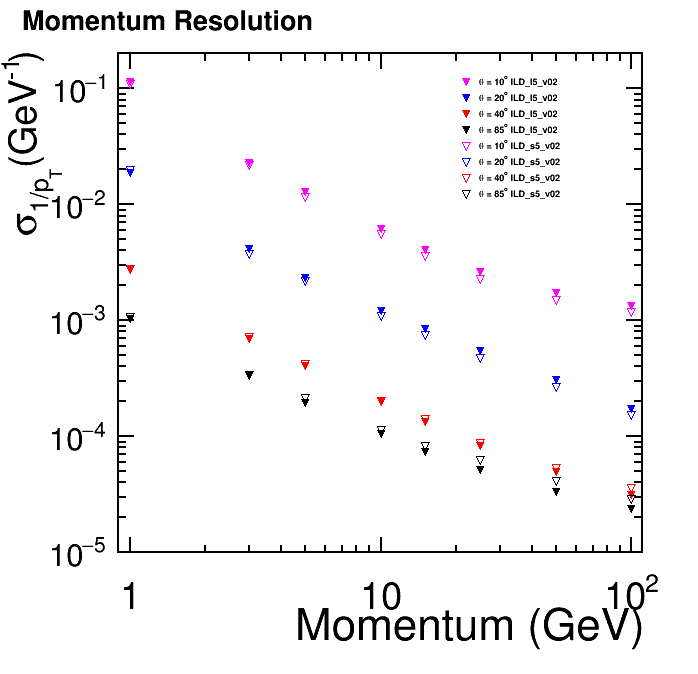}
      }
      \subcaptionbox{}{
         \includegraphics[trim={2.5cm 12.5cm 2.8cm 2.5cm },clip,width=0.45\textwidth]{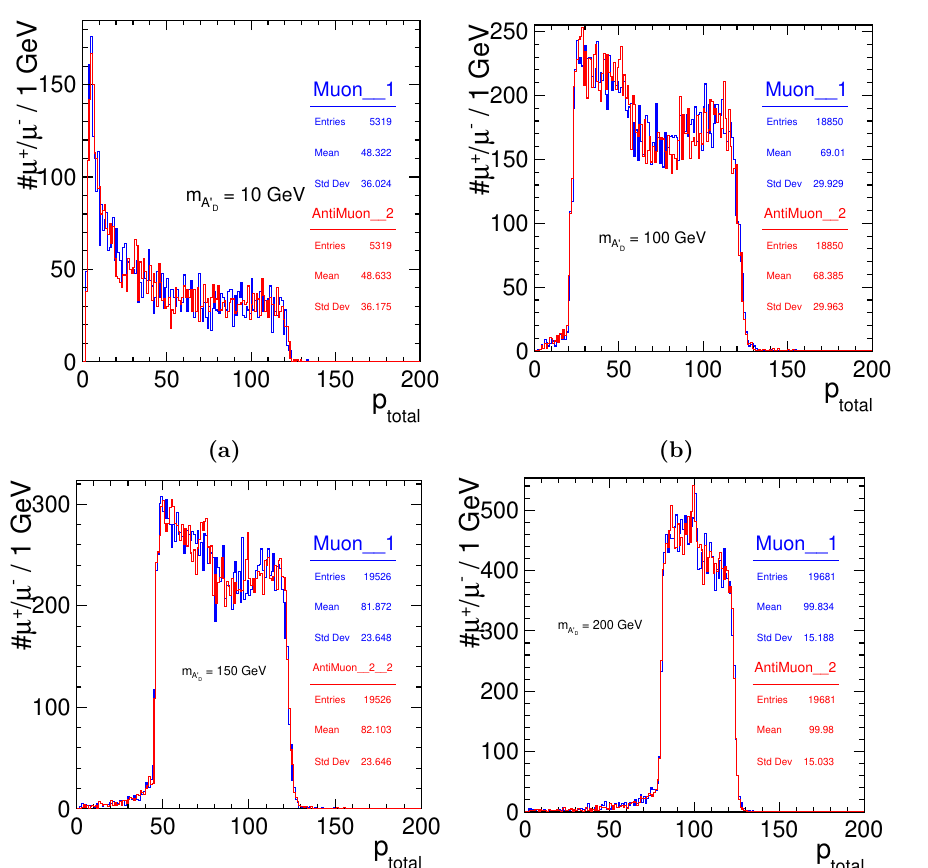}
      }
    \end{center}
    \caption{
     (a): Momentum resolution for charged particles in ILD from full detector simulation;      
     (b): Momentum of muons from  $A_D$ decays at several values of $m_{A_D}$.
\label{fig:sigmapt_and_pmu}}
    \end{figure}

\begin{figure}[b]
    \begin{center}
      \subcaptionbox{}{
        \includegraphics [trim={0cm 0.0cm 0cm 0cm },clip,width=0.5\textwidth]{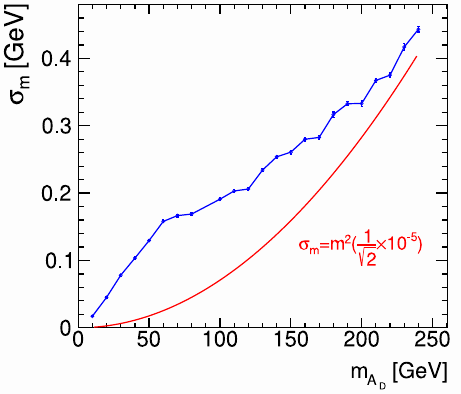}
      }
      \subcaptionbox{}{
        \includegraphics [trim={0cm 0.0cm 0cm 0cm },clip,width=0.45\textwidth]{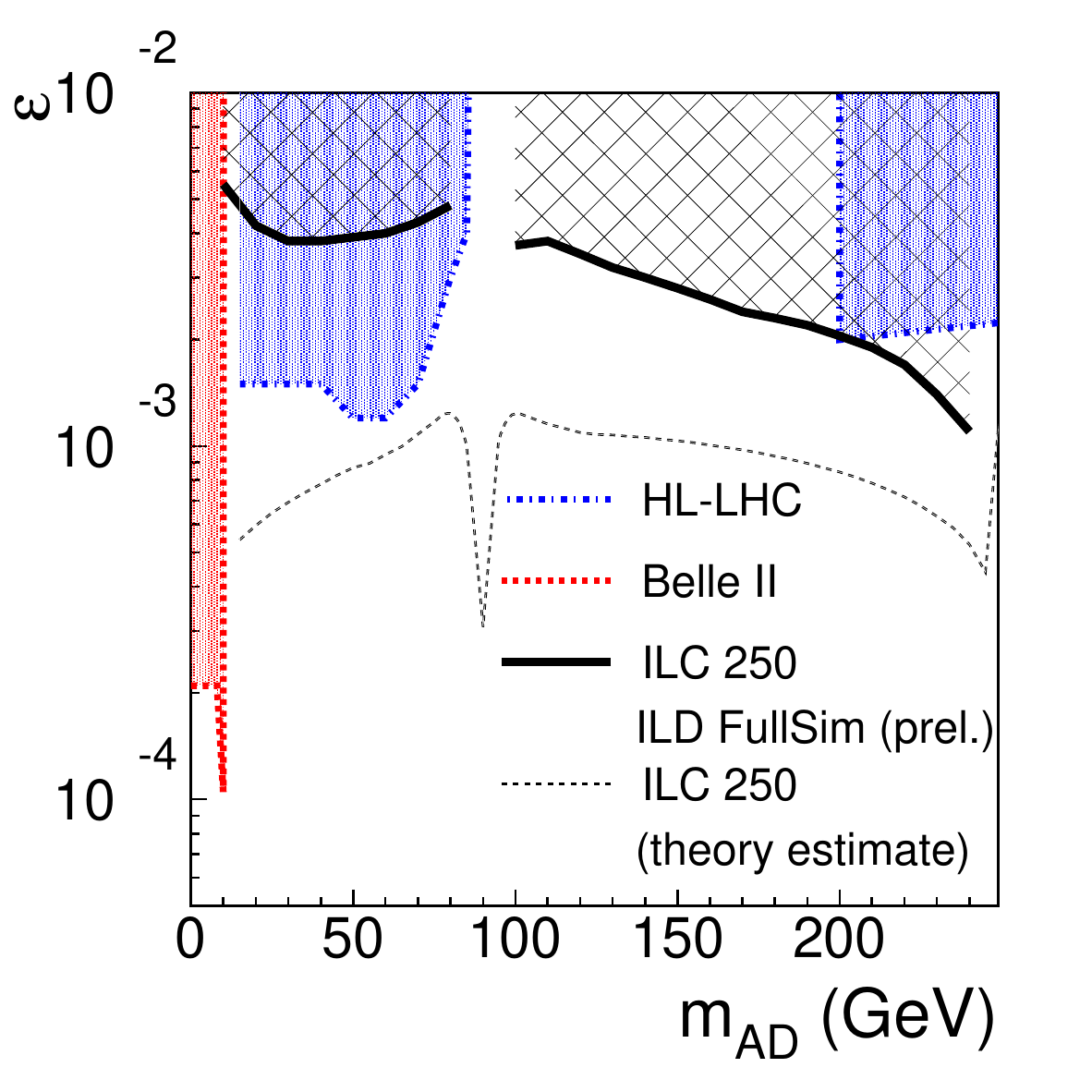} % ilc-250_sc_w_belle_lhc_wkey_fullsim_mod}
      }
    \end{center}
    \caption{(a): The di-muon mass resolution versus $m_{A_D}$. The blue curve is
      the full simulation results, the red one is the simplified theory level one used in
      \cite{Karliner:2015tga,EuropeanStrategyforParticlePhysicsPreparatoryGroup:2019qin};
      (b): The exclusion reach of ILC 250 obtained from this full simulation study of ILD,
      and the expectations of
      Belle II and HL-LHC (from \cite{EuropeanStrategyforParticlePhysicsPreparatoryGroup:2019qin}).\label{fig:exclusions}}
    \end{figure}

One can make a simple estimate
     of the mass resolution
     by assuming that the direction of the ISR lies along the beam axis.
     Under the further assumption that $\sigma(1/p_T)$ is independent of $p$,
     and since $M^2 = p_1p_2(1-\cos \theta_{12})$,
     error propagation
     yields that $\sigma_m \propto m^2$.
     This is the
assumption used in \cite{Karliner:2015tga} to obtain the exclusion limits presented in the 
briefing book (Figure \ref{fig:EPPSUdarkphoton}),
which also assumes that the background consists exclusively of  $\eeto \mu^+\mu^- + \gamma_{ISR}$.
However, due to
multiple scattering, for $p \lesssim 100$~GeV, $\sigma(1/p_T)$ is \textit{not}  constant,
and one observes a strong      % \ne k p^2$ instead of  $\sigma(p) \ne k p^1$ due to the strong
dependence
on $\theta$ when the muon is detected in the forward region rather 
than in the barrel region (Figure \ref{fig:sigmapt_and_pmu}(a)).
Figure \ref{fig:sigmapt_and_pmu}(b) shows that most produced muons have momenta
where multiple scattering dominates the momentum resolution,
and Figure \ref{fig:effmu_and_thetamu}(b) shows that most of the muons are at lower
angles (below the barrel region of the tracker).
Furthermore, the ISR is not always at a zero angle to the beam: if this were the case, the muons would lie exactly on the
curves with the greatest number of entries in Figure \ref{fig:effmu_and_thetamu}(b).
    In reality, none of the assumptions regarding mass resolution - the red curve in Figure \ref{fig:exclusions}(a) - used for
    the EPPSU curve are valid. The correct and complete result of the full simulation
    is the blue curve.
    Due to the considerable fluctuations in momentum resolution as a function of momentum and polar angle
    as well as the angle of the ISR photon,
    the resolution will vary considerably from one event to another.
  An event-by-event simulation is therefore essential.

    Note that since the uncertainty of the mass is known, event-by-event, the search can be optimised
    for sensitivity
    by finding the factors  $f_{low}$ and $f_{high}$ in an event-specific search window from $f_{low}\sigma_m$ to
     $f_{high}\sigma_m$ around the probed mass that yields the highest sensitvity.
     These factors are different above and below the tested mass, because of final state radiation off the
     muons, which tends to produce a tail in the observed mass distribution towards lower masses.

    \section{Expected exclusions at ILC250 and recasts to LCF250, LCF550 and LCF1000}
The result obtained in \cite{Hosseini-Senvan:2024}
with full simulation is shown in Figure \ref{fig:exclusions}(b).
Compared with the simple, theory level, estimate (Fig \ref{fig:EPPSUdarkphoton}(b)), one sees that
at the highest mass, the correct limit
    is a factor two higher than the na\"ive estimate,  a factor four at 100 GeV.
This is due to estimating the momentum uncertainty correctly.
Below $M_Z$, the difference is larger, and in fact the HL-LHC limits
    are expected to be stronger.
Here, the reason is both due to  using a correct error-estimate,
    but also due to the much larger background from non-$Z\rightarrow\mu\mu$
    processes.

The result of the full simulation analyis at ILC250 can be re-cast to quite
acurately estimate the reach at the Linear Collider Facility operating at 250, 550 or 1000 \GeV.    
For LCF250 it is quite straight-forward: one only need to  scale to the higher luminousity expected
at LCF250 w.r.t. ILC250.
One should keep in mind that
    $\sigma \propto \epsilon^2$, so twice the luminosity limit will improve with
the inverse of the \textit{fourth} root of the luminosity increase, not the square root.
Hence, the increase in luminosity from ILC250 to LCF250, i.e. from 2 to 3 ~ab$^{-1}$
only reduces the limit in $\epsilon$ by  $1/\sqrt[4]{1.5}$=0.90.
In addition, Likelihood-weighting of the different polarisation samples can be applied -
this was not done in~\cite{Hosseini-Senvan:2024}.
However, since the signal and the background
from $e^+e^- \rightarrow \mu^+ \mu^-$
have the same polarisation dependence,  Likelihood-weighting  only helps a lower masses,
          where other backgrounds, with different polarisation dependence, dominate.

\begin{figure}[t] 
    \begin{center}
        \includegraphics [width=0.6\textwidth]{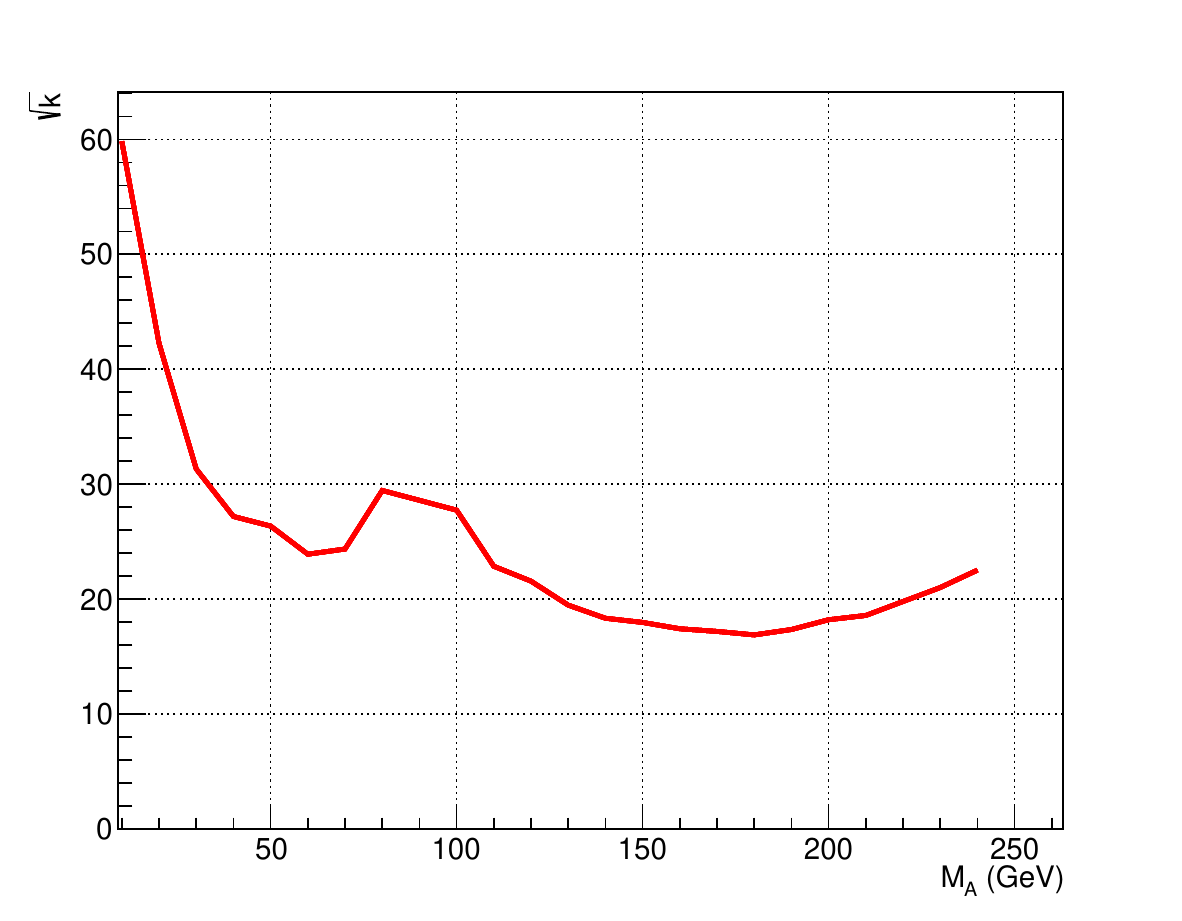}
   \end{center}
\caption{The factor $\sqrt{k}$ relating the signal and background levels
at the 95\% CL bound.}\label{fig:sqrtk}
\end{figure} 
At the higher centre-of-mass energies at LCF550 and LCF1000, several aspects enter.
The luminousties are different, the cross-section for the signal will be different, 
and higher dark photon masses can be probed. 
To take this into account, 
it is suffcient to recalculate the cross-section at different, higher masses and
centre-of-mass energies with the same tools (\textsc{Whizard} and the UFO model) as
were used for the ILC250 analysis.
The branching-ratio to muons
change with the dark photon mass.
However, lower $E_{CM}$ will always give better limits up to the
    kinematic limit, so for LCF550 we can start scanning  $M_A >$  at 250 GeV,
    for LCF1000 at 550 GeV.
For those masses, the branching ratio to muon-pairs is constant
(see Figure \ref{fig:cross_sections_and_width}(b)),
as is  the efficiency to find the muon-pairs (see Figure \ref{fig:effmu_and_thetamu}(a)).
In addition,
the background is found to be entirely composted by $e^+e^- \rightarrow \mu^+\mu^-$
at these higher masses.
The remaining question is  what the background level will be at constant signal efficiency.
This can be evaluated as follows: 
For dark photons to muon-pairs, the signal cross-section is proportional
    to the SM cross-section for $e^+e^- \rightarrow \mu^+\mu^-$.
Therefore, for $M_A >$ 250 GeV (actually for $M_A \gtrsim$ 100 GeV already)  , 
    signal and background scale the same way.
 To keep the same signal efficiency, the search window should
    scale with the mass-resolution ($\sigma_m$).
 At all masses, this will be a quite narrow window, so that the background will be flat.
Therefore,  one expects that the background will scale
    with the mass-resolution,
and this mass-resolution can be calculated
\footnote{This calculation is not trivial,
and will be the topic of a future publication.}.

\begin{figure}[t] 
   \begin{center}
       \subcaptionbox{}{
      \includegraphics [width=0.45\textwidth]{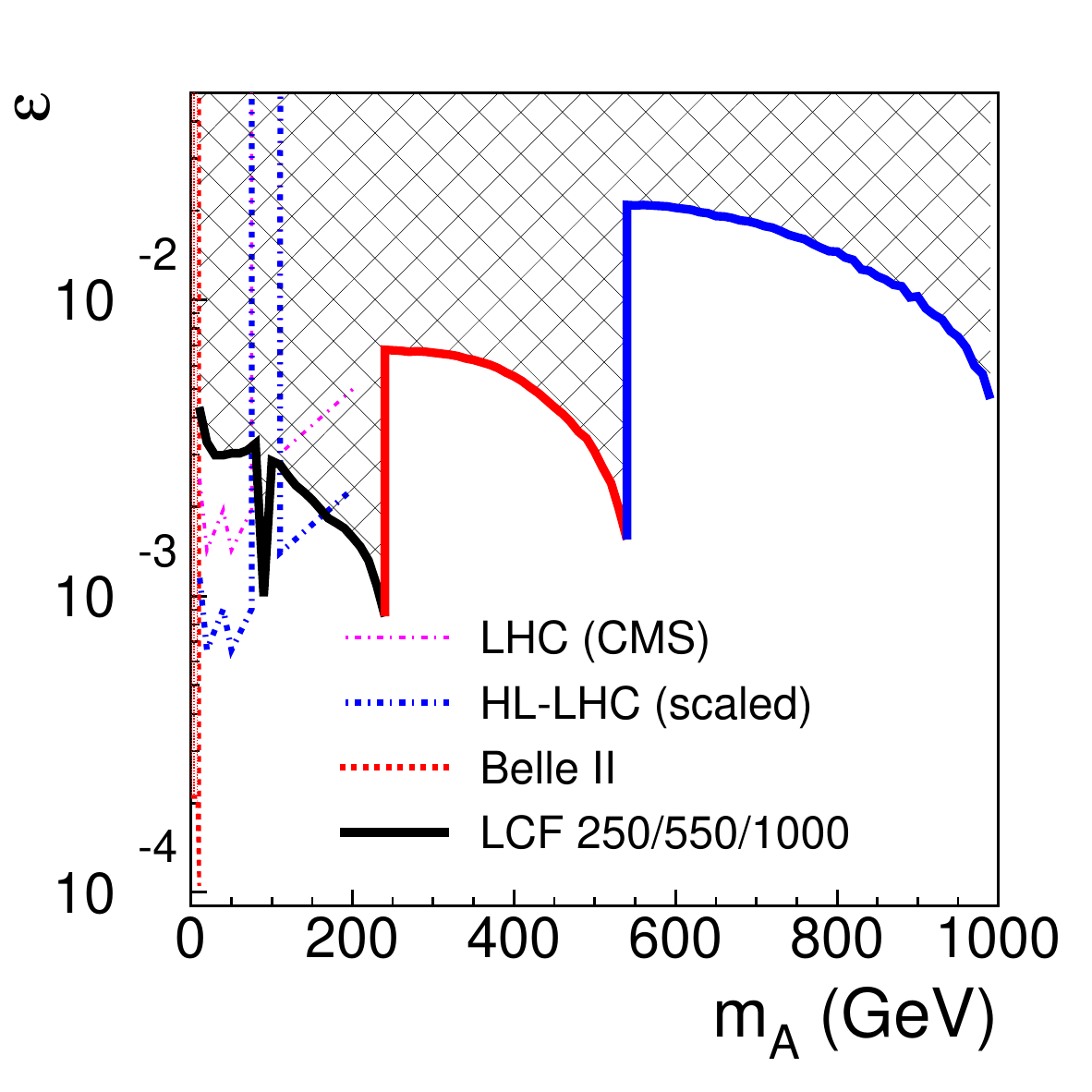}
      }
      \subcaptionbox{}{
      \includegraphics [width=0.45\textwidth]{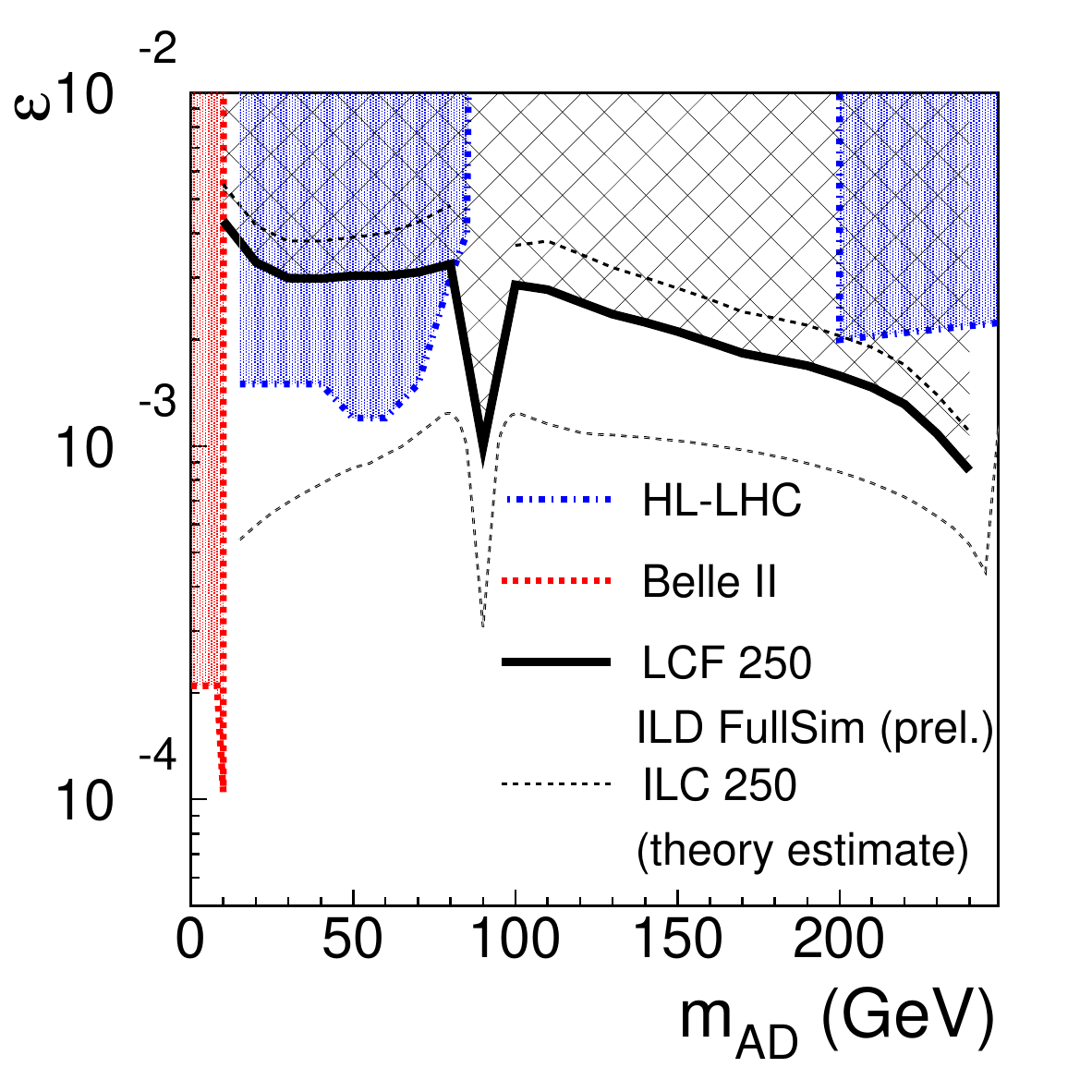}
      }
    \end{center}
\caption{(a): Expected limits at LCF250, LCF550 and LCF1000, after doing
the recast explained in the text. The current limits from CMS and Belle II, as well as a scaling of the CMS limits to HL-LHC are also shown; (b): a zoom to the masses up to 250 GeV, also showing the 
ILC250 full-sime result, used as an input to the LCF recasts. }\label{fig:ilclcf250limandrecasts}
\end{figure}

If one assumes that the number of observed background events ($B$) scales
with  the number of observed signal events ($S$), i.e. that $B=k \sigma_m S$,
one can determine what the factor $k$ is at limiting value of $\epsilon$ by
noting that
        at the limit $S_{lim}/\sqrt{B} = 2$. 
By putting in the numerical values found in the 
          Full-sim analysis at ILC250 one can solve for $\sqrt{k}$ :
          $\sqrt{k}=\frac{1}{2} \frac{1}{\sqrt{\sigma_m}} \frac{1}{\sqrt{\sigma \eta Br \mathcal{L} \epsilon^2_{lim}}}$.
The result is shown in Figure \ref{fig:sqrtk},
and it can be concluded that  $\sqrt{k}\approx $ constant = 20 once $M_A >$~100~GeV.
This argument can then be inverted
 to calculate $\epsilon^\prime_{lim}$ at another point, with different (but known) $\sigma_m , \sigma , \mathcal{L}$ and $\sqrt{k}=20$:
 to yield $\epsilon^\prime_{lim}= 2 \sqrt{k} \sqrt{\sigma^\prime_m} \frac{1}{\sqrt{\sigma^\prime \eta Br \mathcal{L}^\prime }}$.
The resulting exclusion-reach for all LCF options is given in Figure \ref{fig:ilclcf250limandrecasts}(a),
and in Figure \ref{fig:ilclcf250limandrecasts}(b) the case of LCF250 is shown, compared to
the ILC250 curves both for the full simulation analysys, and the theory estimate.

\section{Conclusion and outlook}
 Even in the simplest topologies - or perhaps precisely in those - a full simulation is required
    to obtain a realistic result.
    For in these cases, accuracy is the most important factor.
    However, even if the correctly evaluated range is significantly lower than
    the theoretical estimate, the e$^+$e$^-$ colliders allow the investigation of weaker dark photon 
couplings  than at the HL-LHC, at least for masses greater than $M_Z$.

It would be interesting to have a serious evaluation of the prospects for FCC-ee,
up to its limited maximum reach of 365 GeV.
Since  full detector simulation, with full
SM background is lacking for FCC-ee,
firm conclusions will have to wait.
At lower masses, the huge luminosity at FCC-ee, in particular at the Z-pole, will certainly
out-perform LCF, and might even compete with Belle II.
However, it is probable that the results at FCC-ee at the higher masses will not be  better than LCF: 
FCC-ee will collect more luminosity, but the detectors will necessarily be less performant,
since the FCC-ee conditions require lower detector B-field, and more material due
to the need of cooling.
Hence, the  mass-resolution, crucial for this analysis, will be worse.

   This is an ongoing study. There is still room for improvement in the analysis.
    We have considered only the muon channel. This is because the muon channel
    is expected to exhibit the best mass resolution. 
It would also be possible to include $A_D \rightarrow e^+e^-$;
    however, to achieve sufficient mass resolution,
    it would be necessary to develop a method to compensate for bremsstrahlung.

Furthermore, by utilising the properties of the detected ISR, it should be possible
        to reduce background noise when $m_{A_D}$ is small.
        To make the most of the insights gained regarding errors at the event level,
        one could also consider an approach based on the unbinned maximum likelihood method.

       Finally, it is clear that the sensitivity is greatest when the collider
        operates at or near the dark photon mass. One might therefore consider
        devoting part of the operating time to a scan of $E_{CMS}$, provided that this 
does not pose any difficulties from the point of view of
        the facility's operation. Among the proposed Higgs factories, only the LCF, ILC and C3 offers this 
possibility.

\section{Acknowledgements}
We would like to thank the LCC generator working group and the ILD software
working group for providing the simulation and reconstruction tools and
producing the Monte Carlo samples used in this study.
This work has benefited from computing services provided by the ILC Virtual
Organisation, supported by the national resource providers of the EGI
Federation and the Open Science GRID.

\printbibliography
\end{document}